# Accurate space potential measurements with emissive probe in high pressure plasma and neutral gas

Jian-quan Li[1], Wen-qi Lu[2, *], Jun Xu[2], and Fei Gao[2]

[1]MOE Key Laboratory of Fundamental Physical Quantities Measurement & Hubei Key Laboratory of Gravitation and Quantum Physics, PGMF and School of Physics, Huazhong University of Science and Technology, Wuhan 430074, People's Republic of China

[2]Key Laboratory of Materials Modification by Laser, Ion and Electron Beams, School of Physics, Dalian University of Technology, Dalian 116024, People's Republic of China

[*]Author to whom correspondence should be addressed E-mail: luwenqi@dlut.edu.cn.

## Abstract

The accurate measurement of plasma potentials at high pressure is investigated with the emissive probe. The emissive probe *I-V* characteristic of argon plasma obtained at 100 Pa is interpreted in detail, showing two inflection points in the *I-V* trace. The accurate plasma potentials at high pressure can be determined from the first inflection point potentials (the higher potential) by the improved inflection point method of the emissive probe, while the second inflection point is a result of the additional ionizing phenomenon caused by the collision between the emitted electrons and the neutral argon atoms. Besides, the accuracy of the plasma potential measurement at high pressure is verified to be 0.3 V by the measurement of space potential distribution between two parallel plates in neutral gas at 100 Pa. The accurate measurement of space potentials in high pressure plasma and neutral gas indicates that the improved inflection point method of the emissive probe is promising in the study of collisional sheath structures and the electrostatic probe diagnostic of high pressure plasmas.

## 1. Introduction

The plasma potential is one of the most important parameters in plasma diagnostics serving as a principal factor in analysis of the Langmuir probe *I-V* characteristics for determining other plasma parameters, such as the electron density, the electron temperature and the electron energy distribution function (EEDF).[1,2] For the plasma potential measurements, the emissive probe is considered superior to the Langmuir cold probe with a wide range of application circumstances, such as RF discharge plasmas,[3-6] plasmas with potential fluctuations,[6-10] magnetized plasmas,[2,11,12] plasma sheaths,[2,13-15] tokamaks,[9,16-19] as well as vacuums.[2,20-22] Of the many existing emissive probe techniques, the inflection point method improved by us can obtain the most accurate plasma potentials.[3,12,23-25] Based on the fact that the inflection point potential of the emissive probe *I-V* characteristic changes linearly with the probe heating current, the improved inflection point method is proposed to linearly extrapolate the relation between the inflection point potential and the probe heating current to zero emission for obtaining the accurate plasma potential.[20,23-25] With the realization of an automatic emissive probe apparatus which can automatically execute the conventionally cumbersome procedure of the improved inflection point method, the accuracy of the method in plasma potential measurements has been greatly improved to be 0.1 V, demonstrated by the measurement of vacuum space potential distribution between two parallel plates.[20,24,25]

Although it has been proved that the emissive probe can provide an accurate measurement of the plasma potential at low pressure (≤1.33 Pa), there are few reports about the plasma potential measurement at a pressure more than 1.33 Pa.[26-29] As the pressure increased, the reliable measurement of the plasma potential is more challenging because that the additional ionization of the neutral particles caused by the increase of the collision rate between the emitted electrons and the neutral particles can inevitably perturb the local plasma.[28,30] Different from the emissive probe *I-V* characteristics obtained at low pressure, it has been validated that there are two inflection points in the *I-V* traces obtained at high pressure due to the extra ionization caused by the emitted electrons.[27,28] However, so far the method to obtain the accurate plasma potentials at high pressure with the emissive probe is still controversial, and the mechanism of the ionizing phenomenon during the emissive probe diagnostic of high pressure plasmas also needs further discussion.[26-29]

In this paper, we have studied the emissive probe diagnostic of electron cyclotron resonance (ECR) plasmas with a discharge pressure from 1 Pa to 350 Pa, showing how to obtain the accurate plasma potential at high pressure with the improved inflection point (IIP) method. As the research results show that the extra ionization of the neutral particles during the emissive probe diagnostic only occurs with the plasma pressure greater than 13.3 Pa, we define the plasma discharged with a pressure greater than 13.3 Pa to be high pressure plasmas. In addition, the measurement of the space potential distribution between two parallel plates in neutral gas at 100 Pa suggests that the IIP method is trustworthy in the space potential measurement at high pressure with an excellent accuracy of 0.3 V, showing that the method is promising in the study of collisional sheath structure and the electrostatic probe diagnostic of high pressure plasmas.

## 2. Experimental setup

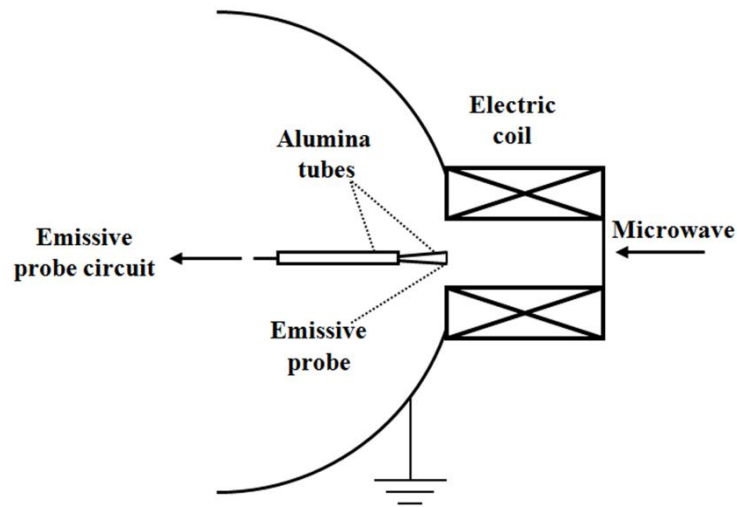

Fig. 1. Schematic diagram of the experimental device for space potential measurements in MW-ECR plasmas with the emissive probe.

The experiments of plasma potential measurements were executed in a microwave electron cyclotron resonance (ECR) plasma setup, as shown in Fig. 1. In the experiments, the chamber was exhausted to a pressure baseline of $1.6\times10^{-3}$ Pa and filled with Ar up to 1 Pa, 10 Pa, 50 Pa, 100 Pa, 150 Pa, 200 Pa, 250 Pa, 300 Pa as well as 350 Pa, respectively. The typical microwave power was 500 W, with the electric coil current of 102 A for ECR discharge. As a thinner emissive probe can achieve a more accurate measurement of the plasma potential,[31] the emissive probe used in the experiments was chosen to be a filament of tungsten wire with 20 μm in diameter and 5 mm in length. As shown in Fig. 2, the filament was wound around two conductive wires (tungsten wires of 0.2 mm in diameter). With folding and compressing the connecting end of the conductive wires, it guaranteed a reliable electrical contact between the probe filament and the conductive wires.[32] In order to minimize the influence of the emissive probe on the plasma, the probe holders (two alumina tubes with about 10 cm in length) were made especially small with inner diameter 0.5 mm and outer diameter 1 mm. Besides, behind the probe holders, it was an alumina tube with double holes for fixing the probe holders and insulating the conductive wires from plasmas, as also shown in Fig. 1.

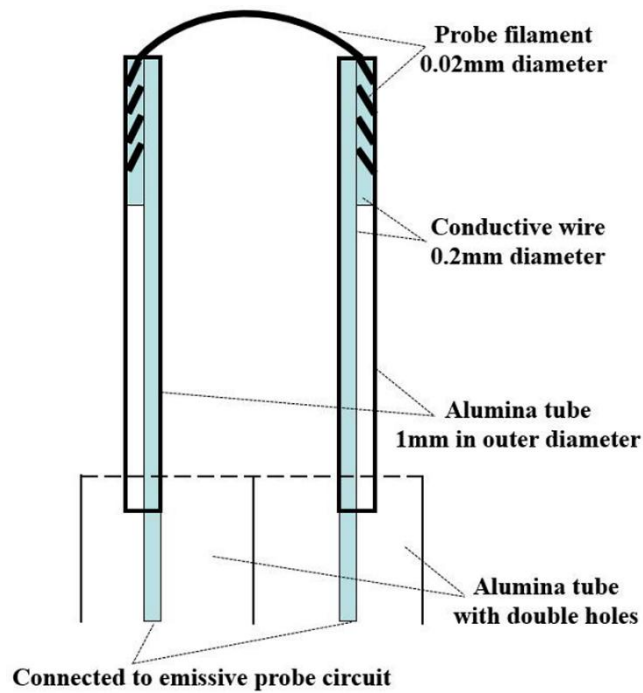

Fig. 2. (Color online) Schematic diagram of construction of the emissive probe.

In the experiment of the space potential distribution measurement, inside the chamber, we placed two stainless steel plates with diameter of 15 cm, in parallel, separated by 5 cm.[20] Between the two plates, a potential difference 10.30 V was applied, supplied by an outside DC voltage source with a potential accuracy of 0.01 V. The potential distribution on the axis of the parallel electrodes calculated by the basic electrics can be considered as the theoretical values. In addition, an emissive probe with the same size (20 μm in diameter, 5 mm in length) was parallel placed between the plates, connected with a driver for moving the probe along the perpendicular center line, as shown in Fig. 3. Before the experiment, in order to prolong the lifetime of the tungsten filament, the chamber was also exhausted to a pressure baseline of $1.6 \times 10^{-3}$ Pa and then filled with argon up to 100 Pa.

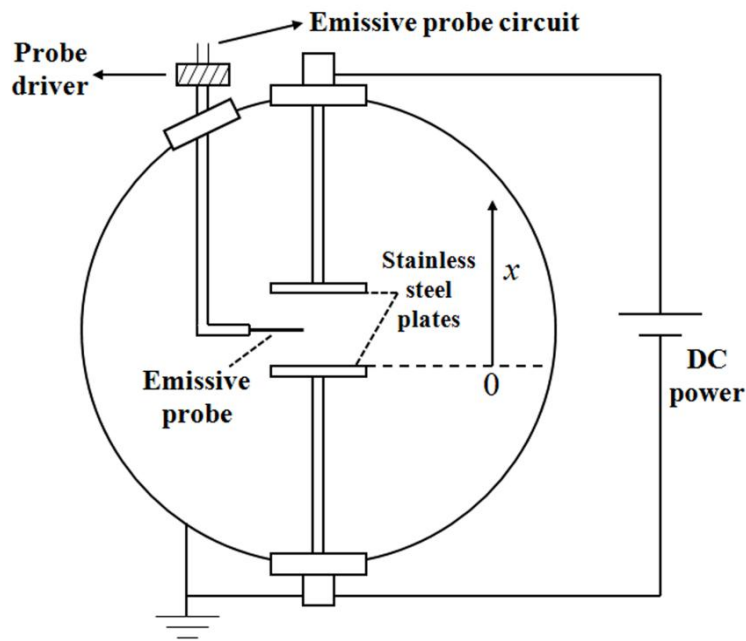

Fig. 3. Schematic diagram of the experimental device for space potential distribution measurements in neutral gas.

In all experiments, the emissive probe was connected with an automatic emissive probe apparatus developed by us, and in the apparatus, the emissive probe *I-V* characteristics were measured point to point, from low potential to high potential, with each point obtained from an average of 500 measurements for suppressing the circuit noises and improving the measurement precision.[20,33] Moreover, in the program of the apparatus, the probe heating current step and the probe scanning bias step were respectively chosen to be 2 mA and 0.2 V.

## 3. Measurement results

### 3.1 Results of plasma potential measurement at 1 Pa

Fig. 4 shows a typical emissive probe *I-V* characteristic and its derivative curve obtained in the ECR argon plasma at 1 Pa, with a clear view showing one inflection point ($V_{ip}$) in the *I-V* trace. In the ideal conditions where the space charge effect is negligible, the inflection point gives the correct plasma potential. However, in practice, the space charge effect caused by the emitted electrons is unnegligible. As the probe bias approaches to $V_p$, the emitted electrons can't be completely absorbed by the plasma. The electrons accumulated around the probe can form a potential barrier which prevents the increase of the emission current ($I_{emis}$), and the potential barrier can shift the inflection point away from the actual plasma potential. With the increase of the probe temperature, the number of electrons accumulated around the probe increases. On the other hand, some high-energy electrons can override the potential barrier and enter into the plasma.[34-36] The interaction between the two factors causes $V_{ip}$ changes almost linearly with the probe filament heating current ($I_{ht}$), as shown in Fig. 5. Therefore, the accurate $V_p$ can be obtained by linear fitting the $V_{ip} - I_{ht}$ curve to the probe heating current at which the electron emission initiates, and this is the improved inflection point (IIP) method proposed by us.[23-25] As shown in Fig. 5, the accurate plasma potential at 1 Pa is determined to be 12.07 V by linearly fitting the $V_{ip} - I_{ht}$ curve to zero emission (140 mA). Besides, the IIP method is also capable in the vacuum space potential measurements.[20]

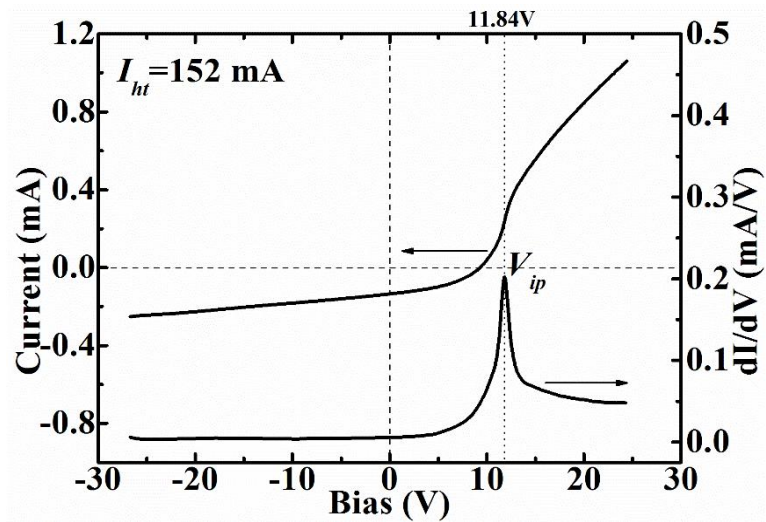

Fig. 4. A typical emissive probe $I$-$V$ characteristic and its derivative curve obtained in argon plasma at 1 Pa.

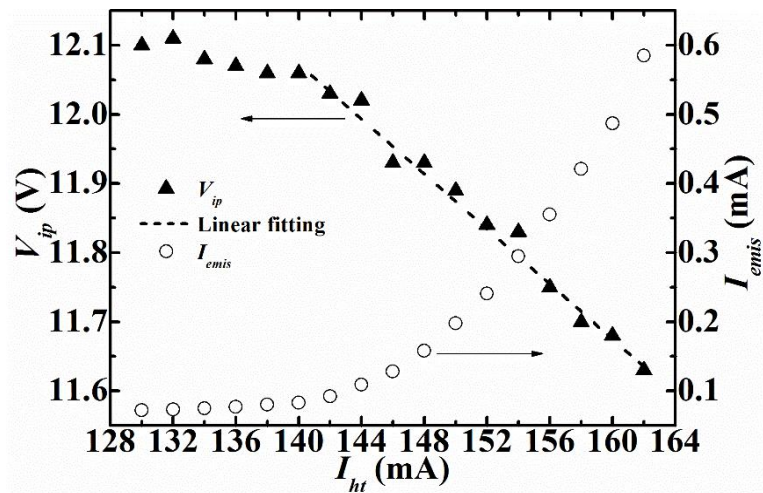

Fig. 5. Changes of $V_{ip}$ and $I_{emis}$ with $I_{ht}$ obtained in argon plasma at 1 Pa.

*3.2 Results of plasma potential measurement at 100 Pa*

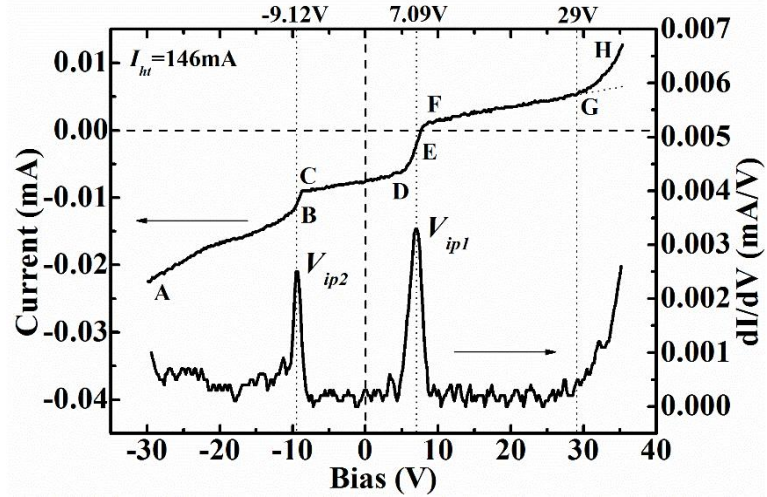

Fig. 6. A typical emissive probe *I-V* characteristic and its derivative curve obtained in argon plasma at 100 Pa.

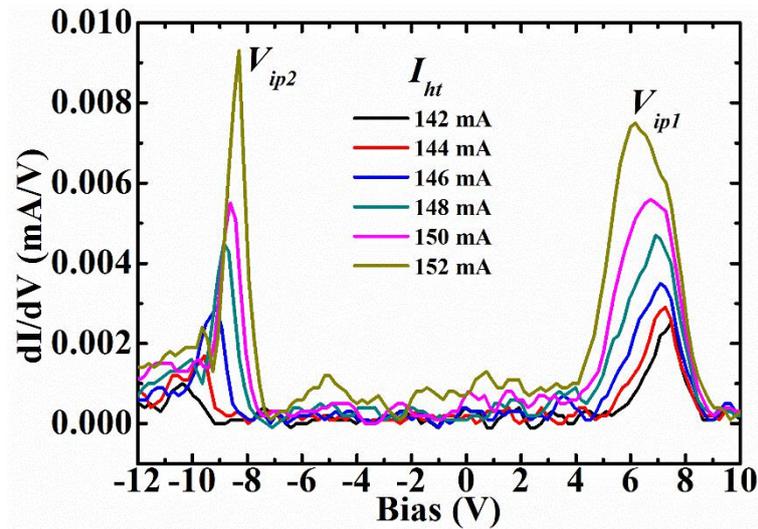

Fig. 7. (Color online) Derivative curves of the emissive probe *I-V* characteristics with different probe heating currents obtained in argon plasma at 100 Pa.

Different from the emissive probe *I-V* characteristic of low pressure plasmas, there are two inflection points (point B and E) in the *I-V* trace obtained in the ECR argon plasma at 100 Pa, as shown in Fig. 6. To distinguish the two inflection points, we define the inflection point with a higher potential to be the first inflection point ($V_{ip1}$), and the other to be the second inflection point ($V_{ip2}$). In order to research the formation mechanism of the two inflection points, we graph the two inflection points with different probe heating currents ($I_{ht}$). As shown in Fig. 7, with the increase of $I_{ht}$, $V_{ip1}$ and $V_{ip2}$ gradually change to be lower and higher, respectively.

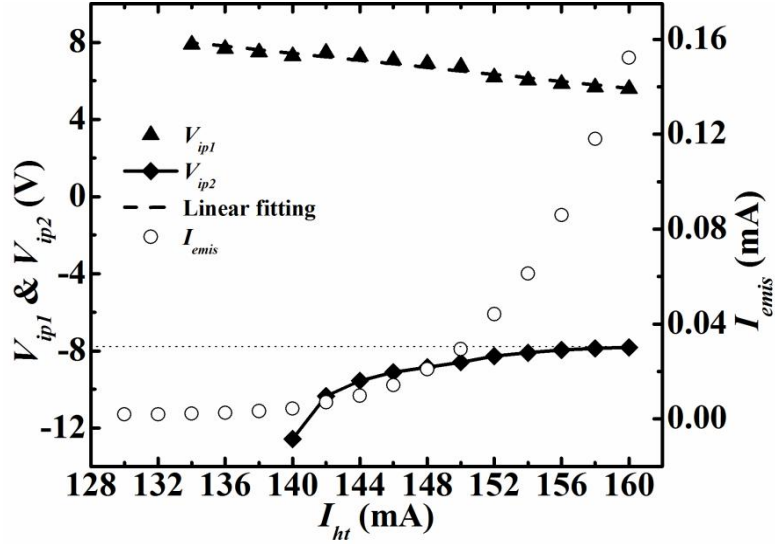

Fig. 8. Changes of $V_{ip1}$, $V_{ip2}$ as well as $I_{emis}$ with $I_{ht}$ obtained in argon plasma at 100 Pa.

Further drawing studies of the two inflection point potentials and $I_{emis}$ with $I_{ht}$ show that $V_{ip1}$ almost decreases linearly with the increase of $I_{ht}$, while $V_{ip2}$ first increases and then becomes saturation, as shown in Fig. 8. Combined the $V_{ip} - I_{ht}$ curve shown in Fig. 5, it is clear that $V_{ip1}$ is caused by the electron emission current which is limited by the space charge. Applying the IIP method to the $V_{ip1} - I_{ht}$ curve, the accurate $V_p$ at 100 Pa is determined to be 7.98 V. On the other hand, Fig. 8 also presents that the saturated potential of $V_{ip2}$ ($V_{ip2-sat}$) is about −7.90 V. The potential difference ($V_d$) between $V_p$ and $V_{ip2-sat}$ is about 15.9 V, approximately equal to the ionization potential of the argon atom (15.8 V), demonstrating that $V_{ip2}$ is caused by the additional ionization of neutral argon atoms. Note that the second inflection point always appears after the emissive probe begins to emit electrons, as shown in Fig. 8. Hence, there is reason to believe that the ionization of argon atom is a result of that the emitted electrons collide with the neutral argon atoms while they are accelerated away from the probe surface. In addition, as the thermoelectron emission from the filament is increasing, the virtual cathode can be formed in front of the filament to self-limit the maximum possible electron emission from the filament.[34-36] When the temperature of emitted electrons is small, the electrons can't override the potential barrier unless an extra electric field to accelerate them. However, when the temperature of emitted electrons is large, some high-energy electrons can override the potential barrier and enter into the plasma,[35] and at this time the ionization of argon atoms occurs as the potential difference between the probe bias and the plasma potential approaches to the ionization potential of the argon atom (15.8 V). That is why $V_{ip2}$ first increases and then becomes saturation with the increase of $I_{ht}$.

Based on a clear understanding of the extra ionization of neutral argon atoms during the emissive probe diagnostic of high pressure plasmas, the *I-V* trace shown in Fig. 6 can be interpreted in detail. The segment FH is the electron collection current. The segment DF and CD mainly represent the space charge limited emissive current and the temperature limited emissive current, respectively. And the segment AC can be identified as the sum of the temperature limited emissive current and the ionization current caused by the collision between the argon atoms and the emitted electrons. Besides, in the program of the automatic emissive probe apparatus, $I_{emis}$ is determined to be the average of the segment CD.

## 3.3 Results of space potential measurement in argon gas at 100 Pa

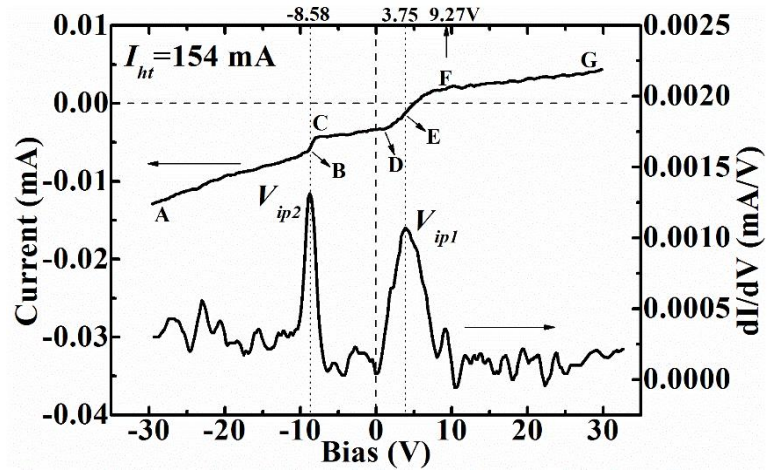

Fig. 9. A typical emissive probe *I-V* characteristic and its derivative curve obtained in argon gas at 100 Pa.

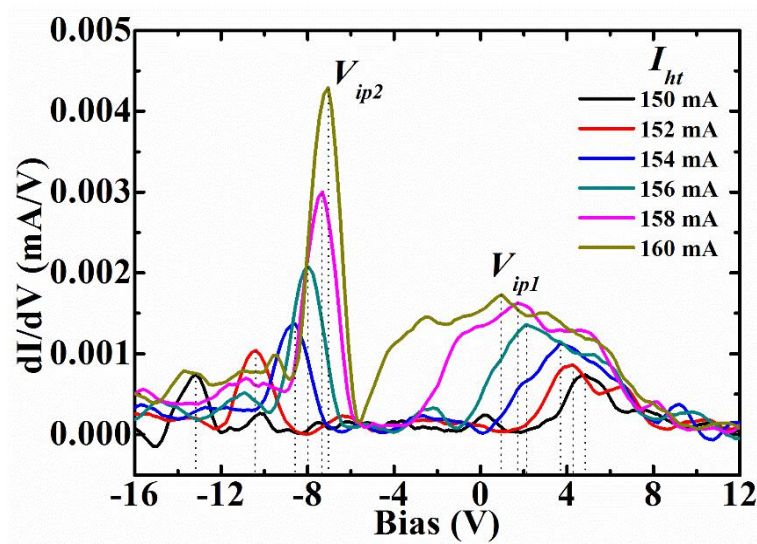

Fig. 10. (Color online) Derivative curves of the emissive probe *I-V* characteristics with different probe heating currents obtained in argon gas at 100 Pa.

Fig. 9 shows a typical emissive probe *I-V* trace and its derivative curve obtained in argon gas at 100 Pa. Similar with the *I-V* trace shown in Fig. 6, there are also two inflection points (point B and E) in the *I-V* curve obtained in argon gas at 100 Pa. As shown in Fig. 10, the relations between the two inflection point potentials and $I_{ht}$ are same with those obtained in the plasma potential measurement at 100 Pa, suggesting that the additional ionization of neutral argon atoms may also exist in the space potential measurement in argon gas at 100 Pa with the emissive probe, and the accurate space potential can be determined from the $V_{ip1} - I_{ht}$ curve.

To obtain the accurate space potential between two parallel plates in argon gas at 100 Pa, we have also investigated the changes of $V_{ip1}$, $V_{ip2}$ as well as $I_{emis}$ with $I_{ht}$, as shown in Fig. 11. Not accidentally, the relations shown in Fig. 11 are very similar with those displayed in Fig. 8. On the one hand, the relation between $V_{ip1}$ and $I_{ht}$ is similar to the change of $V_{ip}$ with $I_{ht}$ which is obtained in vacuum space potential measurements,[20,25] suggesting that the accurate space potential in argon gas at 100 Pa can be determined to be 9.33 V by using the IIP method. The experimental value agrees well with the theoretical value 9.27 V (the emissive probe 4.5 cm from the grounded plate), showing the accuracy of the IIP method in space potential measurements at high pressure. On the other hand, the potential difference ($V_d$) between the theoretical value of the space potential (9.27 V) and the saturated potential of $V_{ip2}$ (about −6.80 V) is about 16.1 V, approximately equal to the ionization potential of the argon atom (15.8 V), demonstrating again that the collision ionization of neutral argon atoms can exist in the space potential measurements at high pressure with the emissive probe.

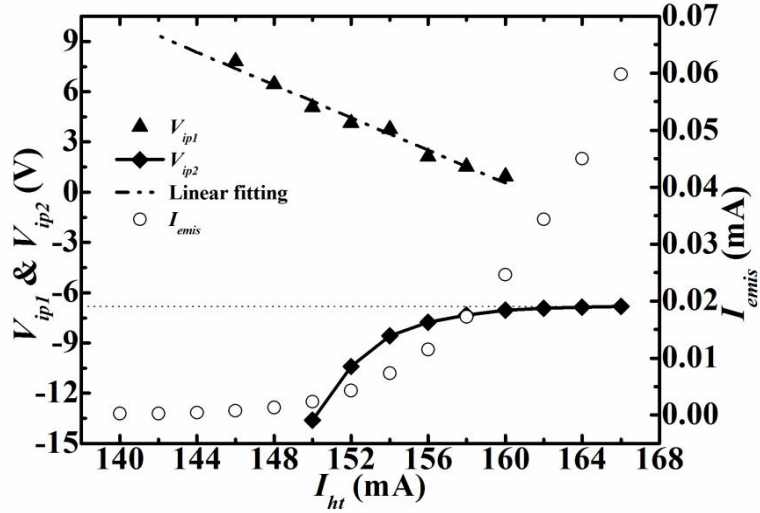

Fig. 11. Changes of $V_{ip1}$, $V_{ip2}$ as well as $I_{emis}$ with $I_{ht}$ obtained in argon gas at 100 Pa.

In addition, compared Fig. 6 with Fig. 9, an obvious difference between the measurement of plasma potentials and space potentials at 100 Pa is that there is a sharp increase in the electron collection current (the segment GH in Fig. 6) of the emissive probe *I-V* characteristics of argon plasma at 100 Pa while it does not appear in the *I-V* curve of argon gas at 100 Pa. The phenomenon can be explained as that it also causes the ionization of neutral argon atoms that the electrons in the plasma collide with the argon atoms while they are collected and accelerated by the probe.

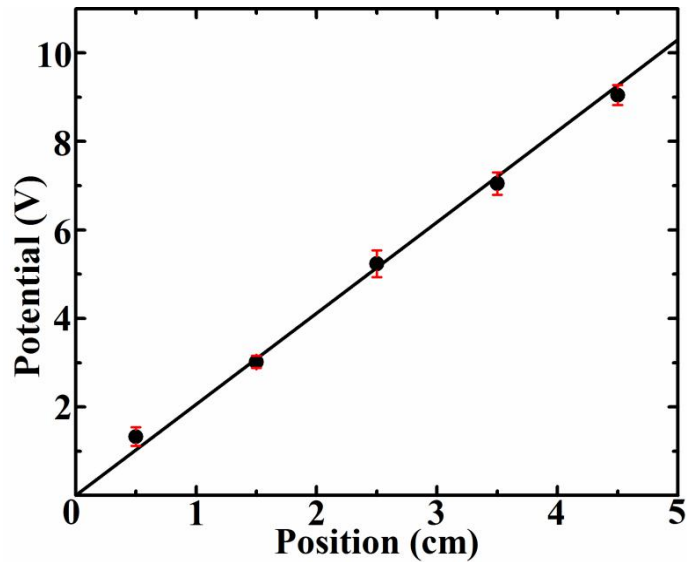

Fig. 12. (Color online) Space potential distribution between two parallel plates measured in argon

gas at 100 Pa.

Using the IIP method of the emissive probe, Fig. 12 shows the space potential distribution determined by the $V_{ip1} - I_{ht}$ curve obtained at 100 Pa. With an average of repeated measurements at different positions between the plates, the measurement results are consistent excellently with the theoretical potential distribution with a measuring error within 0.3 V, suggesting the high reliability and the significant stability of the IIP method in the space potential measurements at high pressure. Besides, with a thoroughly clear understanding of the ionizing phenomenon, the space potential measurements at high pressure with the emissive probe may provide a new approach for the measurement of gas ionization energies.

### 3.4 Results of plasma measurement at different pressure

Based on the space potential measurements in plasma and neutral gas at 100 Pa, we can obtain the accurate plasma potentials at different pressures (from 1 Pa to 350 Pa). The measurement results of plasma potential ($V_p$) and $V_{ip2-sat}$ at different pressures are listed in Table 1. With different pressures, the potential differences ($V_d$) between $V_p$ and $V_{ip2-sat}$ are always about 16 V, suggesting that there is indeed the collision ionization of neutral argon atoms during the emissive probe diagnostic of high pressure argon plasmas.

In addition to the plasma potential, the electron temperature ($T_e$), the electron density ($n_e$) as well as the degree of ionization ($\beta$) are also given in Table 1. With $V_p$ determined accurately by the emissive probe, $T_e$ and $n_e$ can be estimated from the *I-V* curve obtained by a Langmuir probe (a cold emissive probe, 0.02 mm in diameter, 5 mm in length). For example, as shown in Fig. 13, by exponential fitting the current in the transition region of the *I-V* curve, $T_e$ can be calculated using:[28]

$$T_e = e \frac{dV_B}{d\ln(I)} \quad (1)$$

where $V_B$ is the probe bias. And $n_e$ can be calculated from the electron saturation current ($I_{esat}$) which is determined by $V_p$. Besides, $\beta$ is calculated using:

$$\beta = \frac{n_e}{n_e + n_n} \quad (2)$$

where $n_n$ is the neutral particle density which can be obtained from the ideal gas state equation.

Tab. 1. The measurement results of $V_p$, the saturated inflection point potential ($V_{ip2-sat}$), the potential difference ($V_d$) between $V_p$ and $V_{ip2-sat}$, the electron temperature ($T_e$), the electron density ($n_e$) as well as the degree of ionization ($\beta$) at different pressures.

| $P$/Pa | $V_p$/V | $V_{ip2-sat}$/V | $V_d$/V | $T_e$/eV | $n_e$/cm$^{-3}$ | $\beta$ |
|---|---|---|---|---|---|---|
| 1 | 12.18 | None | None | 2.02 | 2.13×10$^{11}$ | 8.79×10$^{-4}$ |
| 10 | 11.22 | None | None | 1.54 | 4.26×10$^9$ | 1.77×10$^{-6}$ |
| 50 | 8.91 | -7.00 | 15.91 | 2.38 | 1.03×10$^9$ | 8.49×10$^{-8}$ |
| 100 | 8.22 | -7.80 | 16.02 | 3.24 | 6.39×10$^8$ | 2.65×10$^{-8}$ |
| 150 | 8.78 | -7.30 | 16.08 | 3.87 | 1.30×10$^8$ | 3.58×10$^{-9}$ |
| 200 | 8.57 | -7.50 | 16.07 | 3.23 | 9.40×10$^7$ | 1.95×10$^{-9}$ |
| 250 | 7.70 | -8.30 | 16.00 | 2.70 | 6.84×10$^7$ | 1.13×10$^{-9}$ |
| 300 | 4.76 | -11.20 | 15.96 | 2.48 | 6.33×10$^7$ | 8.74×10$^{-10}$ |
| 350 | 4.26 | -11.70 | 15.96 | 1.83 | 5.66×10$^7$ | 6.70×10$^{-10}$ |

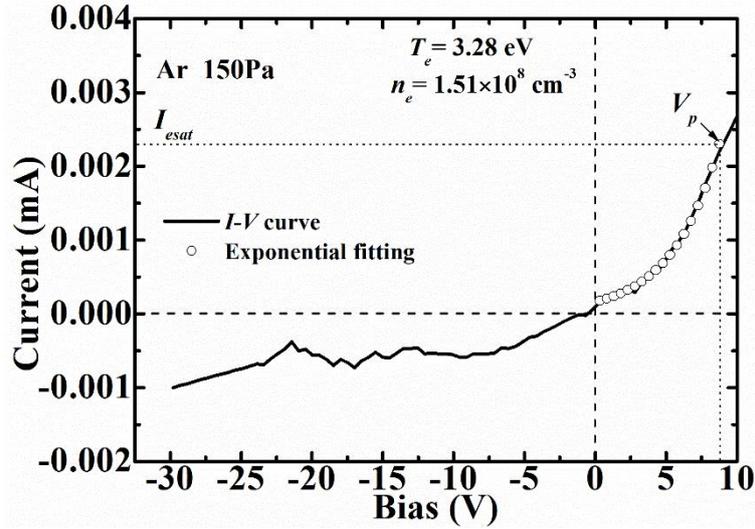

Fig. 13. Langmuir probe *I-V* curve and its exponential fitting curve obtained in argon plasma at 150 Pa.

Besides, according to the theoretical model proposed by Yan *et al.*,[28] the appearance of the second inflection point ($V_{ip2}$) is also closely related to the sheath thickness ($s$) around the emissive probe and the mean free path ($\lambda_i$). The larger $s$ is, the smaller $\lambda_i$ is, and $V_{ip2}$ is more likely to appear. As shown in Table 1, there is no $V_{ip2}$ in the plasma with a discharge pressure under 13.3 Pa, showing the reasons of the high pressure plasma defined by us.

## 4. Conclusions

In conclusion, the measurements of plasma potentials at different pressures (from 1 Pa to 350 Pa) show that there are two inflection points in the emissive probe *I-V* characteristic obtained in the plasma discharged at a pressure more than 13.3 Pa: the first inflection point and the second inflection point. The relation between the first inflection point potentials with the probe heating current is similar with the inflection point obtained in low pressure plasmas, while the second inflection point is a result of the extra ionization caused by the collision between the emitted electrons and the neutral argon atoms. With a clear understanding of the mechanism of the ionizing phenomenon, the accurate plasma potential at a pressure more than 13.3 Pa can be determined from the $V_{ip1} - I_{ht}$ curve by the IIP method of the emissive probe. The accuracy of the IIP method in the plasma potential measurements at high pressure is verified to be 0.3 V by the experiment of space potential distribution measurements between two parallel plates at 100 Pa.

In addition, with the realization of the accurate plasma potential measurements at high pressure, the electron temperature and the electron density at different pressures from 1 Pa to 350 Pa can also be determined by a cold emissive probe (Langmuir probe). The accurate measurement of plasma potentials at high pressure with the emissive probe can provide a foundation for the further research of collisional sheath structure and the electrostatic probe diagnostic of high pressure plasmas, promote the development and application of high pressure plasmas, and may be promising in the measurement of gas ionization energies.

## Acknowledgments

The study was supported by the National Natural Science Foundation of China (No. 91836105 ; No.11675039).